\documentclass{elsart}

\usepackage{graphicx}

\usepackage{amssymb}

\begin{document}

\bibliographystyle{elsart-num}  

\begin{frontmatter}



\title{Time expansion chamber system for characterization of TWIST low
  energy muon beams}


\author[TRIUMF]{J. Hu},
\author[TRIUMF]{G. Sheffer},
\author[TRIUMF,RRC]{Yu.I. Davydov},
\author[TRIUMF]{D.R. Gill},
\author[TRIUMF]{P. Gumplinger},
\author[TRIUMF]{R.S. Henderson},
\author[TRIUMF]{B. Jamieson},
\author[TRIUMF]{C. Lindsay},
\author[TRIUMF]{G.M. Marshall\corauthref{cor}},
\author[TRIUMF]{K. Olchanski},
\author[TRIUMF]{A. Olin},
\author[TRIUMF]{R. Openshaw},
\author[RRC]{V. Selivanov}

\corauth[cor]{Corresponding author. Address: TRIUMF, 4004 Wesbrook
Mall, Vancouver, BC V6T 2A3, Canada. Tel: 604-222-7466. Fax:
604-222-1074. Email: glen.marshall@triumf.ca.}

\address[TRIUMF]{TRIUMF, Vancouver, BC, V6T 2A3, Canada}
\address[RRC]{RRC ``Kurchatov Institute'', Moscow, 123182, Russia}
\
\begin{abstract}

  A low mass time expansion chamber (TEC) has been developed to
  measure distributions of position and angle of the TRIUMF low
  energy surface muon beam used for the TWIST experiment. The
  experiment is a high precision measurement of muon decay and is
  dominated by systematic uncertainties, including the stability,
  reproducibility, and characterization of the beam. The
  distributions measured by two TEC modules are one essential ingredient
  of an accurate simulation of TWIST. The uncertainties, which
  are extracted through comparisons of data and simulation, must
  be known to assess potential systematic uncertainties of the
  TWIST results. The design criteria, construction, alignment,
  calibration, and operation of the TEC system are discussed, including
  experiences from initial beam studies. A brief description of
  the use of TEC data in the TWIST simulation is also included.

\end{abstract}

\end{frontmatter}

\section{Introduction}
\label{introduction}

The TRIUMF Weak Interaction Symmetry Test (TWIST) measures
momentum and angle distributions of positrons from the decay of
highly polarized positive muons, to determine precisely the decay
(``Michel'' \cite{Michel50,Bouchiat57,Kinoshita57}) parameters. It
uses a solenoidal spectrometer with 2 T magnetic field consisting
of 56 planar drift and proportional chambers arranged
symmetrically with high precision on either side of a central
foil that stops the low energy muon beam \cite{Henderson05}.  The
incident polarized muon beam is directed along the axis of the
solenoid from the beam line, through the fringe field region, and
into the chamber stack within the uniform tracking region.
Following the decay of the muon, the positron is tracked in the
chambers and a high statistics decay distribution is acquired.
The first results from the experiment have been published
\cite{Musser05,Gaponenko05}. To extend those results to higher
precision, and especially to analyze and assess systematic
uncertainties for the polarization dependent asymmetry decay
parameter $P_\mu \xi$ \cite{Jamieson06}, a method was required to
measure quickly and reliably the muon beam characteristics near
the entrance to the solenoid. While the TWIST detector itself,
specifically the first several planes of drift chambers, has some
excellent properties for this kind of measurement, its placement
is disadvantageous. It is located after materials that contribute
to multiple scattering, such as a vacuum window and trigger
scintillator. In addition, the gas density at atmospheric
pressure and the extra cathode layers of the first detector
chambers add intolerable scattering sources.

\subsection{Muon beam characteristics}

The requirements of high polarization and rapidly achieved high
statistics are satisfied by a beam of surface muons
\cite{Pifer76} produced at the surface of the primary proton
target from pion decay at rest. High polarization with respect to
the direction of momentum ${\vec{p}}$ ($P_\mu^{\vec{p}}$, which
is exactly $-1$ if the Standard Model holds true) is assured,
because the helicity of the muon neutrino determines the spin
angular momentum of the recoiling muon. The resulting muons have
momentum of 29.79 MeV/c (4.12 MeV) and a range of only
0.15~g~cm$^{-2}$ in carbon.

For the TRIUMF M13 beam line \cite{Oram81}, the primary
production target is typically graphite of thickness 10 mm.  At
the momentum of surface muons, there are also other particles in
the beam, such as positrons, positive pions, protons, and heavier
positive ions. However, the pion decay length is short compared
to the channel length, while protons and heavier ions do not
penetrate thin material layers, leaving positrons as the major
beam contaminant. They can be readily distinguished from surface
muons by a combination of energy loss and time of flight with
respect to the 43~ns period of the accelerator. The time of
flight also permits separation of a small component of muons
which arise from pion decay in flight. While the maximum surface
muon rate for the beam line is of order $10^6$~s$^{-1}$, the
TWIST experiment selects muons to provide a small beam spot,
divergence (see Section \ref{beam_characterization}), and
momentum spread. The muon rate is typically 1--5~$\times
10^3$~s$^{-1}$.

Those muons that are created within the primary target, some
distance from its surface, will lose energy and undergo multiple
scattering prior to escaping into the surrounding vacuum. Some
scattering can also occur in any materials in the beam line
between the primary target and the TWIST detector.  Scattering
reduces the correlation of momentum and spin directions, and thus
the polarization, so it is essential to eliminate such materials
as much as possible and to select only those muons near the
maximum momentum allowed in the decay to keep this depolarization
at a level negligible compared to the sensitivity of the TWIST
measurements.

These criteria demand a muon beam line with momentum resolution
$\Delta p / p \sim 0.01$ (FWHM) \cite{Oram81}, with no
significant windows, residual gas, or other material in the
muons' path. The muons transported to the end of the
beam line by magnetic elements will retain the correlation of
momentum and spin direction, $P_\mu^{\vec{p}}$. The ensemble
polarization of the beam with respect to a beam axis $\hat{z}$,
$P_\mu^{\hat{z}}$, will be less than $P_\mu^{\vec{p}}$ for a beam
of finite emittance, but a high and predictable correlation of
momentum and spin is retained.

As the muon traverses the beam line, the solenoid fringe
field region, and enters the tracking region, the momentum and spin
direction retain their correlation ($P_\mu^{\vec{p}}$ is
preserved), but transverse momentum components are added in the
fringe field and an apparent depolarization with respect to the
solenoid (and beam) axis is the result, i.e., $P_\mu^{\hat{z}}$
decreases.  This can be modeled quite precisely in simulations,
assuming the field shape is known sufficiently well, and also
assuming the particle positions and momenta can be determined for
some point near the entrance to the field.

A detector to measure muon positions and momentum directions in a
surface muon beam has been constructed, using two Time Expansion
Chambers (TECs) \cite{TECref} at low pressure.  A low pressure (4
-- 40 mbar) proportional chamber was tested in \cite{Binon71}.
Detailed investigations of low pressure chambers were described
in \cite{Breskin80}.  A TEC determines the track of an ionizing
particle in one dimension, transverse to the particle's
direction, via the drift time of ionization in a relatively
uniform drift field. The ionization then reaches a region of
higher field where gas amplification takes place near a sense
wire anode. Because beam characteristics are required for both
dimensions ($x$ and $y$) transverse to the beam direction ($z$),
the TWIST TEC system consists of two orthogonal TEC modules.

\section{Design, construction, and operation}

Surface muons are easily scattered, so to minimize multiple
scattering effects, the primary design criterion for the TWIST
TEC system was to maintain low mass. The detector must also
operate in the vacuum of the muon beam channel, which in this
case is not isolated from the primary production target nor from
the accelerator itself. Operation of the detector using low
pressure dimethyl ether (DME) gas allows the use of thin windows
that isolate the chamber gas from the beam line vacuum.  DME,
having a long drift time, small Lorentz angle, and high primary
ionization, is convenient because it is also used for TWIST drift
chambers.  An automated control system ensures that this low
differential pressure is maintained. The overall length of the
TEC system was kept to a minimum compatible with a reasonable
track length necessary for an accurate direction measurement. The
beam size can range up to a few centimeters, so a somewhat larger
active area was chosen, within which the only materials in the
beam path are the entrance and exit windows, the low pressure
chamber gas, and very thin field cage wires required to define
the drift field. The device must be capable of recording muon
rates up to several thousand per second, while its operation
should also not be compromised by a comparable rate of beam
positrons.

Even though the TEC material in the beam path has been minimized,
it scatters the beam too much to be used simultaneously during
TWIST's highest precision muon decay polarization measurements.
The total systematic uncertainty grows as the depolarization
increases, so the TEC system is removed for most muon decay data
taking. Instead, it is used periodically to tune and measure
characteristics and to verify muon beam stability between
beginning and end of data sets taking several days to collect. In
this way, possible muon beam changes during the data set
collection become evident, which may be correlated to other
carefully monitored quantities. Therefore, another important
design feature was that the detector can be installed into and
removed from the beam line quickly, conveniently, and
reproducibly.

\subsection{Design of the TEC system}

\begin{figure}[tbp]
  \begin{center}
    \includegraphics[angle=90,width=150mm]{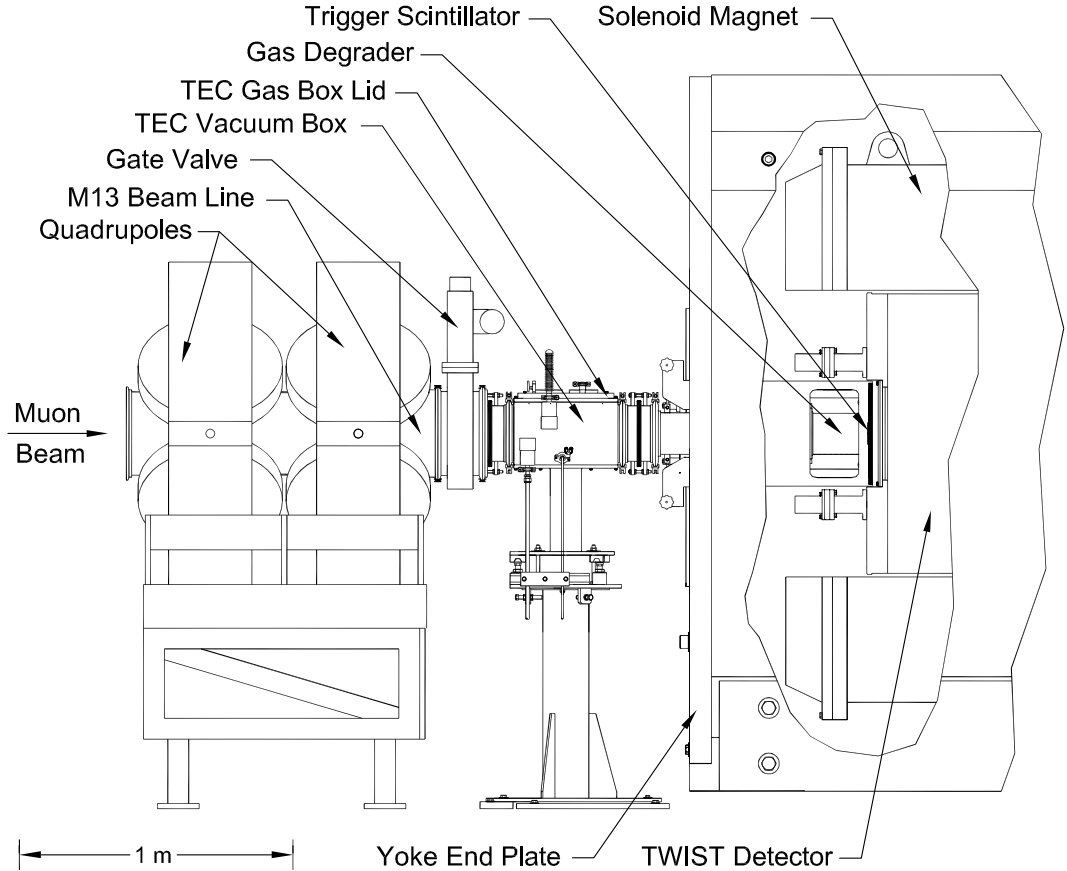}
    \caption{Placement of the TEC system in the muon beam between the
      exit of the beam line and the TWIST solenoid.}
    \label{tec_in_beam}
  \end{center}
\end{figure}

The TEC system is located directly upstream of the TWIST spectrometer
and is attached to the end of the TRIUMF M13 beam line
\cite{Oram81} as shown in Fig. \ref{tec_in_beam}. The beam passes
from the final two quadrupoles through a gate valve and into the
low pressure gas containment volume in which the two
TEC modules operate. It then continues in vacuum through the fringe field near
the yoke end plate until it reaches a vacuum window inside the
high field region. From there, it passes through a variable
density gas degrader, a thin trigger scintillator, and into the
TWIST detector.

\begin{figure}[hbtp]
  \begin{center}
\includegraphics[width=150mm]{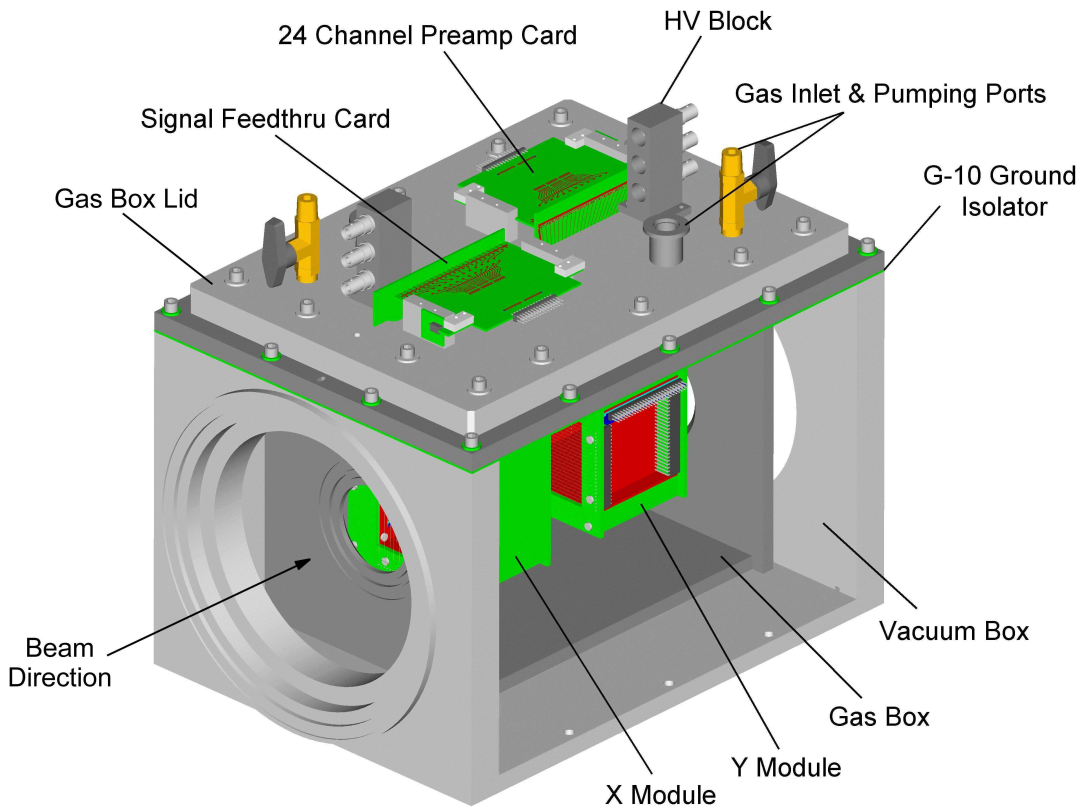}
    \caption{TEC modules in the gas box.}
    \label{two_modules}
  \end{center}
\end{figure}

As can be seen in Fig. \ref{two_modules}, the TEC system consists of two
identical modules rotated 90 degrees with respect to each other
and mounted in an aluminum gas box of length 280~mm.  One module tracks the
horizontal ($x$) position and angle of the incoming muons and the
other the corresponding vertical ($y$) components.  Each module
is 80~mm long. The active area that can be measured by each TEC is
60~mm $\times$ 60~mm transverse to the beam, with an active
length of 48~mm (24~wires spaced by 2~mm).  The gas box has
6~$\mu$m aluminized Mylar entrance and exit windows and is nested
in a vacuum box. The pressure is maintained at 80~mbar with a
flow of DME gas of approximately 100~cm$^3$/min. The vacuum box
is part of the beam line and remains in place, while the TECs,
the gas box, and the gas box lid, can be removed and replaced by
a blank cover plate.

\begin{figure}[hbtp]
  \begin{center}
    \includegraphics[width=150mm]{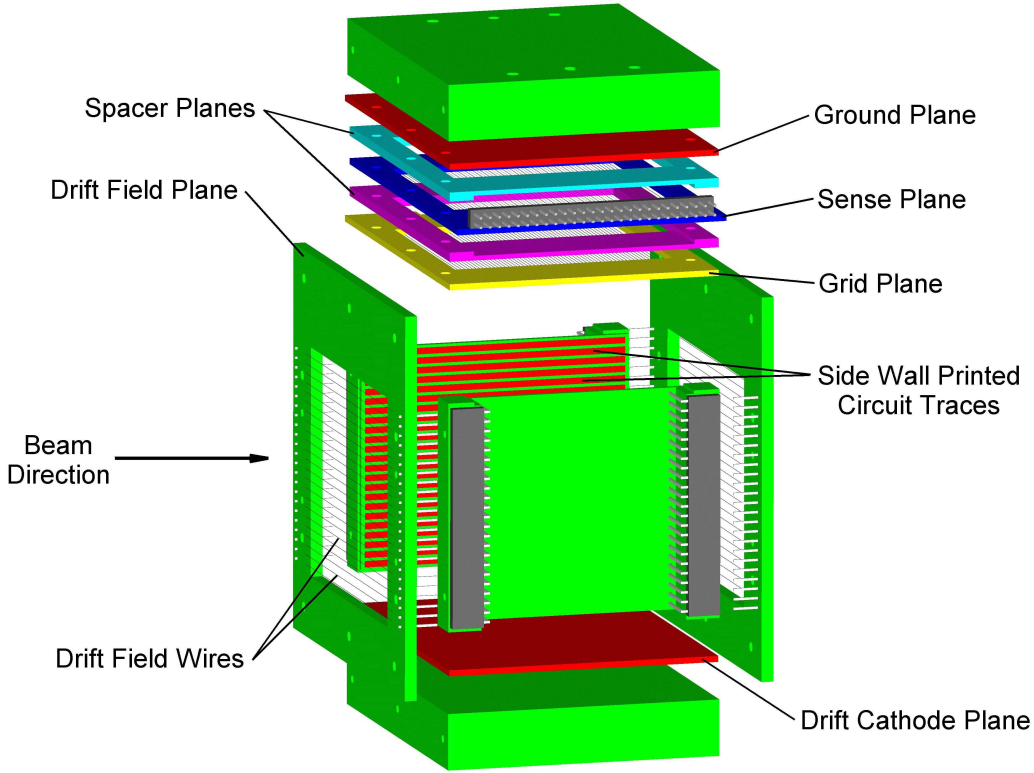}
    \caption{Exploded three-dimensional view of one TEC module.}
    \label{threedim_view}
  \end{center}
\end{figure}

Figure \ref{threedim_view} shows an exploded three-dimensional
rendering of one TEC module. The sense and grid planes are strung
on frames of 
nominally 1.6 mm thick G-10 (epoxy/glass) printed circuit boards
(PCBs) as are the side walls of the field defining cage. The
drift field plane frames are 3.2 mm thick PCBs. The remaining
structural elements of the module are also manufactured from
G-10. The connectors have gold plated contacts in a glass-filled
polyester matrix.  The module is assembled with nylon cap screws.
All materials have been tested for compatibility with the DME gas
\cite{Openshaw03}.

Each module contains a drift field region and gas amplification
region. The drift field region is bounded by a four sided field defining cage
with two side wall elements containing parallel horizontal printed circuit traces
and two end (particle entrance and exit) field planes with 50 $\mu$m gold plated Cu/Be wires
strung horizontally across the opening. The wires of the two end
drift field planes are
electrically connected to the side wall traces through gold
plated pins and a socket assembly.  The pitch of the wires and
traces is 2.54 mm. To achieve a uniform electric field, each
group of wires and traces is connected to its neighbor through
resistors of the same value creating a resistor chain to which
high voltage is applied. The drift cathode plane on the bottom of
the field cage is a copper foil on a 1.6 mm G-10 substrate.

\begin{figure}[hbtp]
  \begin{center}
    \includegraphics[angle=90,width=75mm]{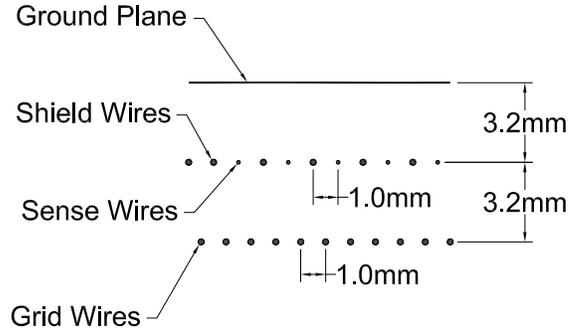}
    \caption{Sense plane geometry.}
    \label{sense_plane}
  \end{center}
\end{figure}

The gas amplification region consists of a grid wire plane, a
sense wire plane and a solid cathode plane. These planes are
separated by spacer planes made of 1.6 mm G-10. A section of the
wire geometry of this configuration is shown in Fig.
\ref{sense_plane}.  The grid and shield wires are 125 $\mu$m gold
plated Cu/Be. The sense wires are 25 $\mu$m gold plated tungsten.
The signals from the 24 sense wires are taken from the end of the
printed circuit board through a coaxial cable assembly to a feedthrough in the gas
box lid.

When the elements of the drift field region and gas amplification
region are mated, the last resistor in the field defining cage is
connected to the grid plane, which is maintained at ground
potential as is the ground plane on the other side of the sense
plane. As shown in Fig. \ref{sense_plane}, the sense plane is made
up of alternating sense and shield wires.  Gas amplification occurs only
at the thinner sense wires. The diameter of 25 $\mu$m was chosen
for these wires based on our previous study of gas gain at low
pressure \cite{Davydov05}.  The shield wires screen the induced
pulses from adjacent wires and are necessary because the average
angle of the muon tracks through the TEC module is quite small. This
results in signals on adjacent sense wires occurring at very
nearly the same time, so the mutually induced pulses would
seriously affect the rise time of these signals and hence the
accuracy of the drift time measurement.

\subsection{TEC Alignment}

It is important to align the TEC system accurately with respect
to the TWIST spectrometer and hence with the measured magnetic
field map.  Cross-hairs were mounted in the apertures at each end
of the steel yoke of the spectrometer to define the $z$ axis. An
optical surveying instrument (theodolite) was then used to align
each TEC to these cross-hairs. First the $x$ and $y$ positions of
the TEC system were adjusted to align the field cage wires for
both $x$ and $y$ modules, and then the location was checked with
the position of the central hole of the calibration collimators
(see Sec.  \ref{extraction_of_str}). The accuracy was
approximately $\pm$250~$\mu$m. Once this alignment was completed
another optical device, a Leica TDA5005 Total Station, was used
to record the position of the TEC vacuum box with respect to the
M13 beam line and TWIST spectrometer.

As described previously, the TECs, the TEC gas box, and the gas
box lid, have been designed to be easily removed and replaced.
This is done in a reproducible manner using locating pins in
the vacuum box, which remains as part of the beam line vacuum
system with a blank cover plate.  However, any time the TEC
assembly is removed and replaced, a measurement of its position
is made with the Total Station to confirm that proper alignment
has been maintained.The accuracy of this measurement is
$\pm$50~$\mu$m. It was necessary to take precautions to avoid
movement due to forces from vacuum loading of the beam line and
magnetic fields of the solenoid.

\subsection{TEC operation}
\label{tec_operation}

The optimum drift field for the TECs was found to be
approximately 16~V/mm, corresponding to an applied drift cathode
plane voltage of -1000~V with the grid plane at ground potential
(refer to Fig. \ref{threedim_view}).
The applied sense-wire high voltage is
1150~V and the shield wires are at 300~V. These voltages result
in a muon track efficiency of essentially 100\% with typically up
to 19 out of 24 wires providing measurable ionization signals (hits) per
muon track (see \ref{tracking_resolution}). There are
intermittent sparks, the location and cause of which are still
undetermined. A degradation of efficiency over a period of weeks
of operation has also been observed, which may be a result of the
sparking. Although the origins of these effects are not yet
understood, the practical solution of regular replacement of
sense wire planes has been adopted.

\subsection{Electronics and readout}

The 24 wires of each of the TEC modules are connected to 24
channel preamplifiers mounted on the lid of the TEC gas box. The
preamplifiers were developed at Fermilab for use at their
Colliding Detector Facility \cite{Yarema92}. Each channel has a
gain of 1 mV/fC and a dynamic range of -400 fC to +20 fC.
Signals are taken via 9.5 m of micro-coaxial cable to custom made
post-amplifier/discriminator modules, which have sixteen channels
in a single width CAMAC module and are also used in the readout
of the TWIST spectrometer \cite{Henderson05}. The resulting
discriminated ECL logic signals are then sent via 15 m of twisted
pair cable to separate channels of a 64-channel LeCroy 1877
FASTBUS TDC (time-to-digital converter). These TDCs are multihit
type to allow digitization of up to eight time intervals per
channel per readout (or event). They have 0.5 ns resolution and
are operated in common stop mode. The trigger signal for the TDCs
is derived from a plastic scintillator mounted downstream of the
TEC vacuum box. When the solenoid is in operation, it is the same
scintillator that provides the trigger for TWIST (as shown in
Fig. \ref{tec_in_beam}); a dedicated trigger scintillator
following a window at the exit of the TEC vacuum box is used for
tests and tuning with the solenoid magnetic field off.

\subsection{Gas system}

The TEC gas system is required to provide a flow (typically 100
cm$^3$/min) of DME to the TECs while maintaining a constant
pressure (typically 80 mbar) in the gas box.  Upstream pressure
control is accomplished with an absolute pressure transducer that
measures the pressure in the TEC volume.  This transducer signal
is fed to a proportional-integral-differential (PID) device
controlling the flow through a mass flow controller in the gas
supply line to the TECs.  Downstream flow control is provided by
a manual fifteen-turn metering needle valve in the
exhaust line between the TECs and the exhaust pump.  The observed
pressure control stability, as reported by the pressure
transducer, is approximately $\pm 0.1\%$.  The manufacturer's
specifications for the pressure transducer indicate an absolute
accuracy (combined linearity, hysteresis and non-repeatability)
of $\pm 0.25\%$ and a temperature drift of $-0.048\% / ^{\circ}$C
at the typical operating pressure. Issues pertaining to TEC gas
box installation, gas control, and interlock features, are
discussed in Appendix \ref{gas_appendix}.


\section{Calibration and Tracking}

\subsection{Field distortion}

The field cage is used to produce a uniform drift field for
electrons. Ideally, the electric field between the drift
cathode plane and the grid plane (Fig.~\ref{threedim_view}) is
perpendicular to the planes and constant at
$E =-1000/60.96 \approx -16.4$ V/mm everywhere in the drift volume.
In reality, however, the field distribution is distorted by the
high voltage applied to the sense plane and also by the stray electric
field from the other TEC module.

\begin{figure}[hbtp]
   \begin{center}
     \includegraphics[width=75mm]{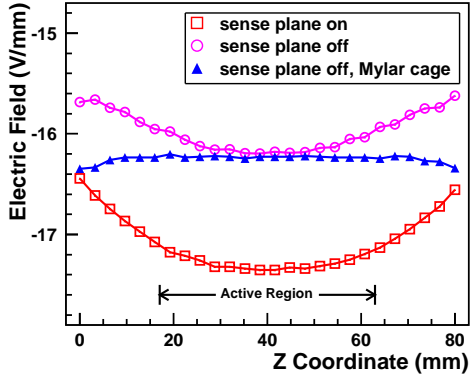}
     \caption{Calculated drift field at the center ($x = y = 0$)
       of a single module. 
      Sense wire plane on and
     off are shown to demonstrate the effect on the dominant field. The
     active region, where the sense wires are located, is from $z = 17$ mm
     to $z = 63$ mm.}
     \label{sense_eff}
   \end{center}
\end{figure}

The FEMLAB Program \cite{femlab} was employed to study such effects.
Figure ~\ref{sense_eff}, obtained from the FEMLAB simulation of a
single module, shows distributions of the drift fields in the center
of the drift volume (i.e., at $x = 0$ for the X module) at various
settings. Compared with the case when the sense plane is off
(open circles), the field changes by about 10\% over the length
of the module when the sense plane is turned on (open squares),
which indicates that the grid plane is not completely effective in
shielding the drift region from the sense plane potential.
A $z$ position dependency of the field intensity is clearly seen
in both cases. This edge effect is likely due to the influence
of the field cage. In a test with FEMLAB, the grid plane was extended
to cover the gap between the edge of the grid plane and the field
cage, and the field cage wires were replaced with aluminized Mylar
strips. The edge effects are then reduced significantly. With the
sense plane off, the drift field is nearly constant across the
module (filled triangles). Because the aluminized Mylar strips
introduce more multiple scattering into the beam, they were not
used in the actual TEC modules. Instead, field variations were accounted
for via the position-dependent calibration procedure described in
the next section.

\begin{figure}[hbtp]
   \begin{center}
     \includegraphics[width=75mm]{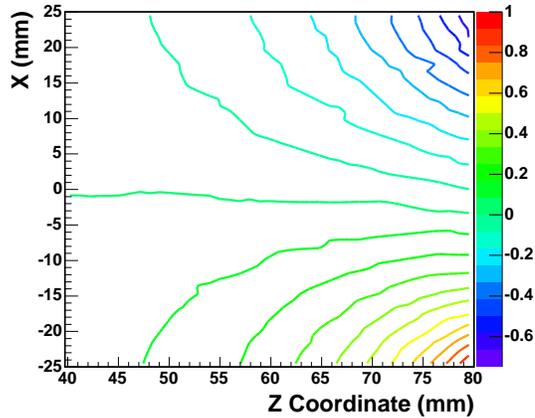}
     \caption{Contours of calculated electric field contribution
      in the X module at $y = 0$, due to
       field leakage from the Y module. Color-coded contours
       show the field change in percentage (right hand color
       scale) of the nominal central field
       (16.4 V/mm) of the X module.
     The difference in the $x$ component of the field, when the Y
     module is turned on, is plotted in the portion of the drift
     region of the X module nearest the Y module. The $z$
     coordinates on this scale for the field cage
     of the Y module are $z = 120$--$200$ mm.}
     \label{module_int}
   \end{center}
\end{figure}

Another source of field distortion comes from interference
between the two modules. Figure~\ref{module_int} is a contour
plot of the change of field in the drift region of the X module
when the Y module is turned on. This could bias the beam angle
measurements by up to roughly 5 mrad even though the interference
effect is on the order of 1\%.

\subsection{Extraction of space-time relationship}
\label{extraction_of_str}

\begin{figure}[hbtp]
   \begin{center}
     \includegraphics[angle=90,width=75mm]{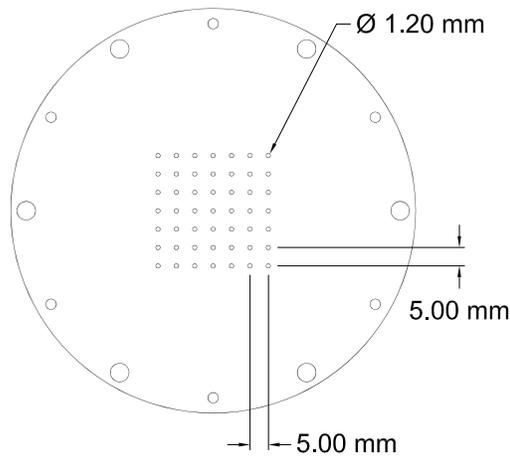}
     \caption{A schematic plot of an aperture array. One is placed at
       each end of the TEC gas box to define tracks to be used
       for calibration.}
     \label{colli}
   \end{center}
\end{figure}

\begin{figure}[hbtp]
   \begin{center}
     \includegraphics[angle=90,angle=-90,width=75mm]{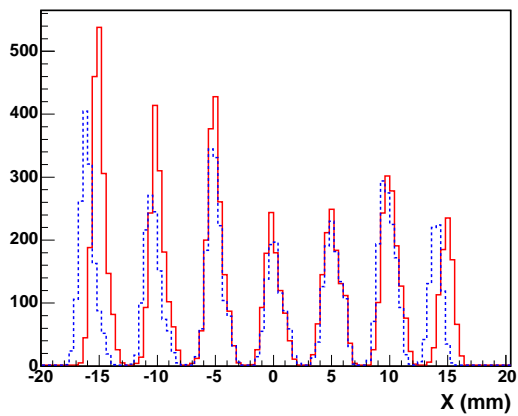}
     \caption{Projection in $x$ of measurements for muons passing
       through holes in one row of the collimator prior to
     calibration (blue dotted line, based on GARFIELD without field
     leakage) and following calibration (solid red line).}  
          \label{array_projection}
   \end{center}
\end{figure}

Precise characterization of the muon beam requires an adequate
STR (space-time relationship, providing the relation between the
distance of a muon track to the sense wires and the transit time
to those wires of electrons produced by ionization along the
track) for a TEC cell, which is defined by two neighboring shield
and grid wires with a sense wire in the middle.  As described in
the previous section, the electric field in the drift region is
not uniform. The drift velocity varies with the drift distance
and the $z$ position of the sense wire. A pair of multi-hole
apertures, installed 141 mm upstream and downstream with respect
to the center of the TEC gas box, acts as a collimator and is
used as an external reference system to measure the STR as a
function of various operating parameters. Each aperture, as shown
in Fig.~\ref{colli}, consists of a square array of $7 \times 7$
holes, 1.2 mm in diameter, separated by 5.00 mm. The hole
positions were measured to an accuracy better than 40 $\mu$m.
Calibration data were obtained with a diffuse beam tune so that
all holes have exposure from the beam. Beam tracks were selected
with an angle near zero mrad for calibration purposes, and it was
verified that these tracks passed through corresponding holes in
each of the two multi-hole apertures of the collimator. An
example of a projection of measured $x$ positions from one row of
the aperture array is shown in Fig.  \ref{array_projection},
before and after the calibration procedure.

\begin{figure}[hbtp]
   \begin{center}
     \includegraphics[width=75mm]{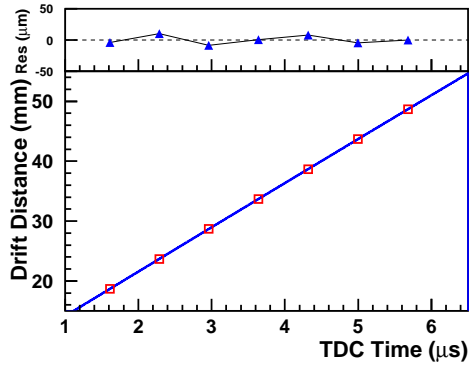}
     \caption{Fit of the drift distance to the uncorrected TDC
       time to derive the STR. The plot at the top of the figure gives the residual of
     the fit.}
     \label{fitstr}
   \end{center}
\end{figure}

A third order polynomial function was used to fit the STR:
\begin{eqnarray*}
   d^k_{ij} &=& p_0^k + p_1^k * t^k_{ij} + p_2^k * (t^k_{ij})^2 + p_3^k *
   (t^k_{ij})^3  \\ & & (k = 1, \cdots, 48, ~{\rm and}~ i,j = 1, \cdots, 7)
 \end{eqnarray*} 
where $d^k_{ij}$ is the distance in $x$ or $y$ between the $k^{th}$ sense wire
and a straight track between corresponding holes in the $i^{th}$ row and the $j^{th}$ column of the
two multi-hole apertures;
$t^k_{ij}$ is the corresponding mean drift time.
Figure~\ref{fitstr} presents an example of the fit quality. The
cubic term reduces residuals by 10--20\%; higher order terms
were found to be negligible.  With $d$ in cm and $t$ in $\mu$s,
the linear term is of order one, while the quadratic and cubic
terms are of order $10^{-2}$ and $10^{-3}$ respectively.  The
drift times for ionization electrons in the active volume of the TEC are
up to $8\mu$s.

While the calibration procedure has been shown to be quite
adequate for the central active area of the TECs ($30 \times
30$~mm$^2$) corresponding to the collimator described above, it
is likely that the derived STR is not as accurate at the
extremities. Apertures that cover the entire active region will
be used in the future.

\begin{figure}[hbtp]
   \begin{center}
     \includegraphics[width=75mm]{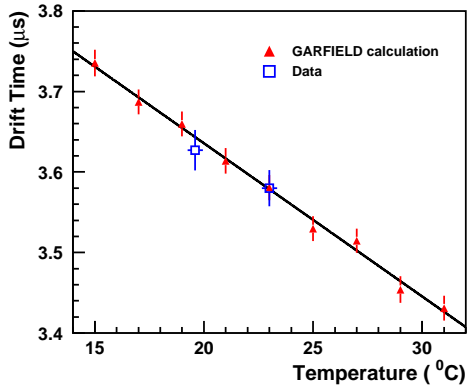}
     \caption{Drift time variation with DME gas temperature. 
The plotted drift times are calculated with GARFIELD at a field of 16.4 V/mm
     for a muon track on the TEC central axis, for various gas
     temperatures.}
     \label{temperature_eff}
   \end{center}
\end{figure}

Simulations with GARFIELD \cite{Garfield} suggest that the drift
time has an almost linear relationship with temperature and
therefore gas density (as illustrated at a field of 16.4 V/mm in
Fig.~\ref{temperature_eff}). Data taken at 19.6$^\circ$C and
23.0$^\circ$C confirm the temperature dependence.  Because the DME
gas pressure is fixed at 80 mbar and the temperature is recorded
by the data acquisition system, variations of the STR with the
gas density have been taken into account either by taking the
calibration data at the same temperature or by linearly
rescaling the STR according to the density.

\subsection{Tracking resolution}
\label{tracking_resolution}

In this paper, we use the word ``event'' to refer to the
information which is recorded for each occurrence of a valid
trigger. The trigger for readout of an event in the TEC system is
provided by a signal from a single plastic scintillator placed
after it. Associated with the trigger within a time window of
$-6$ to $+10$ $\mu$s, one, or rarely more than one, muon may pass
through the TECs. Ionization detected above the electronics
threshold by a sense wire within the $16\ \mu$s time interval
will be referred to as a ``hit''. Minimum ionizing positrons are
also present in the beam, but are detected only with very low
efficiency in the low density gas. In fact, even muons above 50
MeV/c are not tracked efficiently with 80 mbar pressure.
Regardless of the track origin, events with multiple signals in
the trigger scintillator, in the event time window, are rejected
from the analysis. In the track fitting code, track candidate
signals are identified from all wires having signals (hits)
within the time window. The drift time is determined with respect
to the trigger scintillator time, and then converted to drift
distance using the STR.  The drift distance, or transverse track
position, is then fit versus the $z$ position with a straight
line. A straight line is found to be a good approximation even
though the solenoid magnetic field extends to the TEC position,
where it is typically about 0.1 T and mostly in the $z$
direction.

\begin{figure}[hbtp]
   \begin{center}
     \includegraphics[width=75mm]{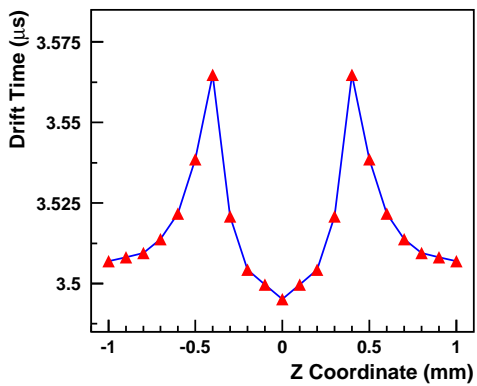}
     \caption{Dependence of the drift time on $z$, from a
       GARFIELD study with drift field 16.4 V/mm 
       for a muon track on the central axis. $z = 0$ mm corresponds to the
     location of the sense wire and $z = \pm 1$ mm to the adjacent
     shield wires, while grid wires are at $z = \pm 0.5$ mm (see Fig.~\ref{sense_plane}).}
     \label{tctspread}
   \end{center}
\end{figure}

There are three major contributions to the resolution of the
chamber: the spread of drift times in a TEC cell, the multiple
scattering effect, and diffusion of electrons.  Due
to the TEC geometry, drift times within the same cell can vary
significantly.  Figure~\ref{tctspread}, obtained from a GARFIELD
study for drift field of 16.4 V/mm, shows the variation of drift time
versus $z$ position of ionization. A variation of up to 2\% can
be seen for an average drift distance, equivalent to 75~ns or
about 0.6~mm.

\begin{figure}[hbtp]
   \begin{center}
     \includegraphics[width=75mm]{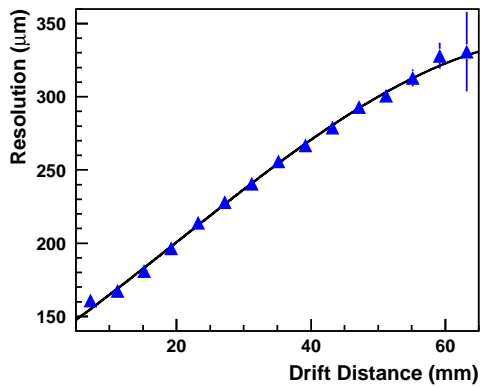}
     \caption{Measured resolution as a function of track distance from
     the sense wire.}
     \label{dres}
   \end{center}
\end{figure}

The spatial resolution of an individual TEC cell is determined by
fitting the tracks without including hits from that cell and then
measuring the residuals. It ranges from 150 to 350 $\mu$m, as a
function of distance from the sense wire, getting worse with
increasing drift distance as shown in Fig.~\ref{dres}.  The
dependence of hit resolution on drift distance could be caused by
transverse and longitudinal diffusion of ionization electrons.
The longer the drift, the less likely is the ionization signal to
exeed the electronic threshold, due to the fact that diffusion
separates the electrons in time.

On average, when there are signals on all 18 sense wires in a
track, the track angle resolution is $\sim 3$ mrad while the
position resolution, when extrapolated to a plane midway between
the two TEC modules, is $\sim 150\ \mu$m.

\begin{figure}[hbtp]
   \begin{center}
     \includegraphics[width=75mm]{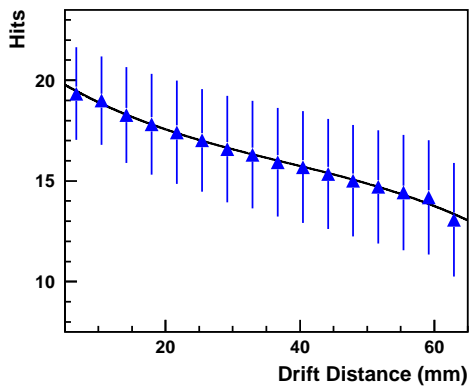}
     \caption{Total number of valid hits as a function of
       track distance from the sense wire. Error bars represent
       rms widths of the hit number distribution.}
     \label{nres}
   \end{center}
\end{figure}

As discussed in Sec. \ref{tec_operation}, while the tracking
efficiency is essentially 100\%, the individual sense wire efficiency is not.
Figure~\ref{nres} is a measured distribution of the total number
of wires with hits (of a maximum possible of 24 wires in a module) for
a track as a function of drift distance, where the inefficiency
due to the longer drift is clearly visible.  Except for this
distance correlation, the absence of a hit on a wire seems to
occur randomly along the length of the track.

\section{Characterization of the TWIST beam}
\label{beam_characterization}

After alignment and calibration of the TECs, it is possible to
obtain and analyze distributions of positions and angles of beam
muon tracks, and to measure correlations between them, in a few
minutes. A characterization of the muon beam with statistical
precision adequate for TWIST simulations can be obtained in a few
hours. This section describes typical measurements and how they
are used by the TWIST group to adjust the beam and to analyze
muon decay data.

\subsection{TEC measurements and implications for TWIST}

The objectives for beam quality for TWIST were established early
in the planning of the experiment. Simulations of a muon beam
entering the solenoidal field showed that, to achieve the
required small depolarization of order $10^{-3}$, the beam should
have an rms size of about 5 mm and rms angle of about 15 mrad in
the absence of the solenoid field, at a position near the field
fringe where radial field components are highest when the
solenoid is on. Therefore, the calibration and resolution
uncertainties of the TECs must be small compared to these values.

Extensive beam studies based on TEC data without the solenoid
field resulted in beam characteristics that satisfied the
requirements. The measurement time necessary for a precise
measure of means and widths of size and angle distributions, as
well as indications of asymmetry or unusual beam shape, was
merely minutes. This allowed many different beam tests and
adjustments.  The effects of slits, apertures, and absorbers in
the M13 beam were observed and settings were optimized.
Quadrupole steering was minimized, and beam stability over days
and weeks was tested.  Two surface muon production targets were
used in the studies. The first was an edge-cooled graphite
surface muon production target of 10 mm length. The second was a
beryllium target of 12 mm length encased in a stainless steel
water cooling jacket; it produced an asymmetric $x$ profile some
20\% larger than the graphite, thus it is considered inferior for
TWIST muon decay measurements.

\begin{figure}[hbtp]
   \begin{center}
     \includegraphics[width=75mm]{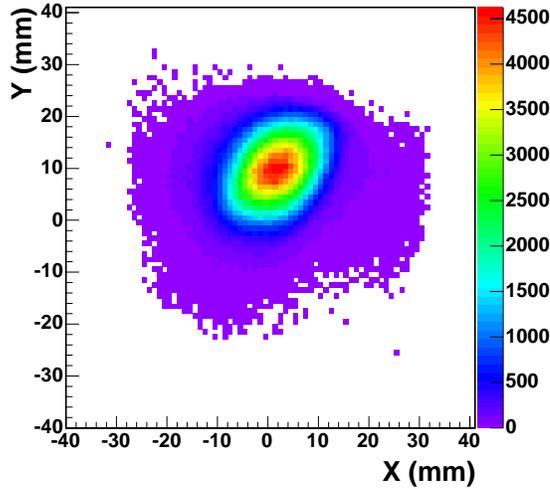}
     \caption{Two-dimensional distribution of the muon beam
       intensity at
       the TEC system, near the
     fringe field region of the TWIST spectrometer.The pixel size
   in the figure is $1 \times 1$~mm$^2$.}
     \label{beam_yvsx}
   \end{center}
\end{figure}

When the solenoid was turned on to its full 2 T central field,
and the STRs from the TEC calibrations were verified to be
consistent with those derived with the solenoid off, it was observed
that the incoming muon beam was steered to positive values of $x$
and $y$. Figure \ref{beam_yvsx} shows an $xy$ beam image with
these conditions, from a graphite production target.  Note that
the intrinsic beam size, without the TEC system, is somewhat smaller
than implied by the figure, because some of the multiple
scattering produced by the TECs occurs prior to tracking in the
TEC modules.  While the solenoid fringe field does not
appreciably alter the beam size, the shift of position away from
the solenoid axis, believed to be due to the interaction of the
solenoid fringe field with beam quadrupoles and the final beam
dipole, is a cause for concern. It reduces the polarization with
respect to the solenoid axis ($P_\mu^{\hat{z}}$), such that it is
less likely to be simulated with the high absolute precision
required by TWIST. Steering in the beam line to compensate for
the fringe field effect is being implemented for future TWIST
measurements.

\subsection{Use of TEC data in TWIST simulations}

The Monte Carlo simulation of each TWIST event begins with the
generation of an input muon. Other beam particles (muons and
positrons) are also generated, randomly correlated in time
according to measured rates, to simulate the real beam
environment.  The muon beam distribution in position and angle
that is input to this simulation is derived from the TEC measured
particle tracks extrapolated to a $z$ position corresponding to
a plane between the TEC modules. At this location, multiple scattering has
an adequately small effect on the transverse ($xy$) profile. The
angular distributions then contain most of the influence of
scattering, and can be modified to account for it to an acceptable level
of precision by a quadratic subtraction procedure. This is
effectively a deconvolution of the intrinsic beam distribution
and the additional broadening of the measured distribution due to
scattering, and is carried out in the simulation.

\begin{figure}[hbtp]
\begin{minipage}[t]{75mm}
  \begin{center}
     \includegraphics[width=\linewidth]{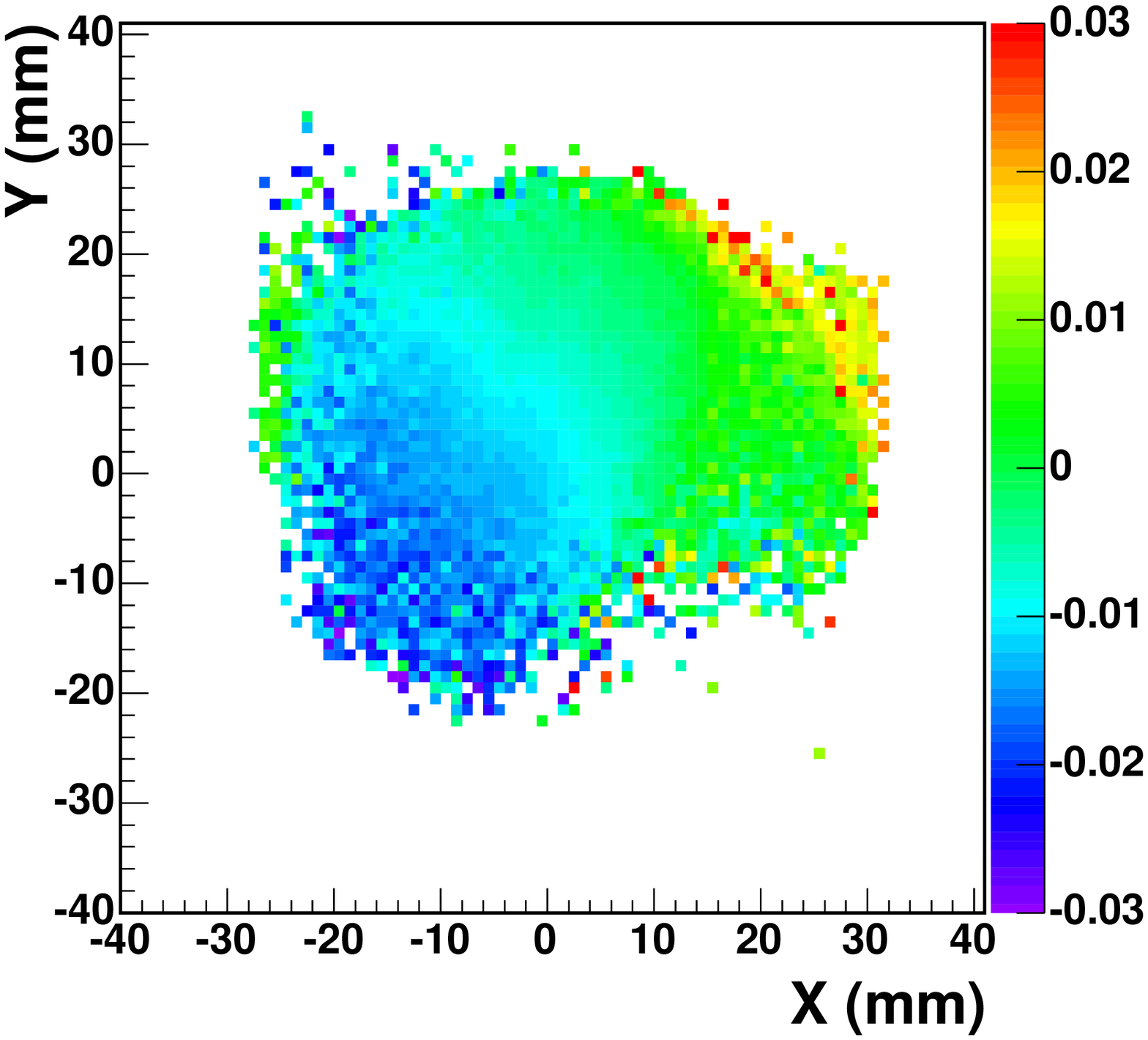}
\end{center}
\end{minipage}
\hspace{2mm}%
\begin{minipage}[t]{75mm}
\begin{center}
\includegraphics[width=\linewidth]{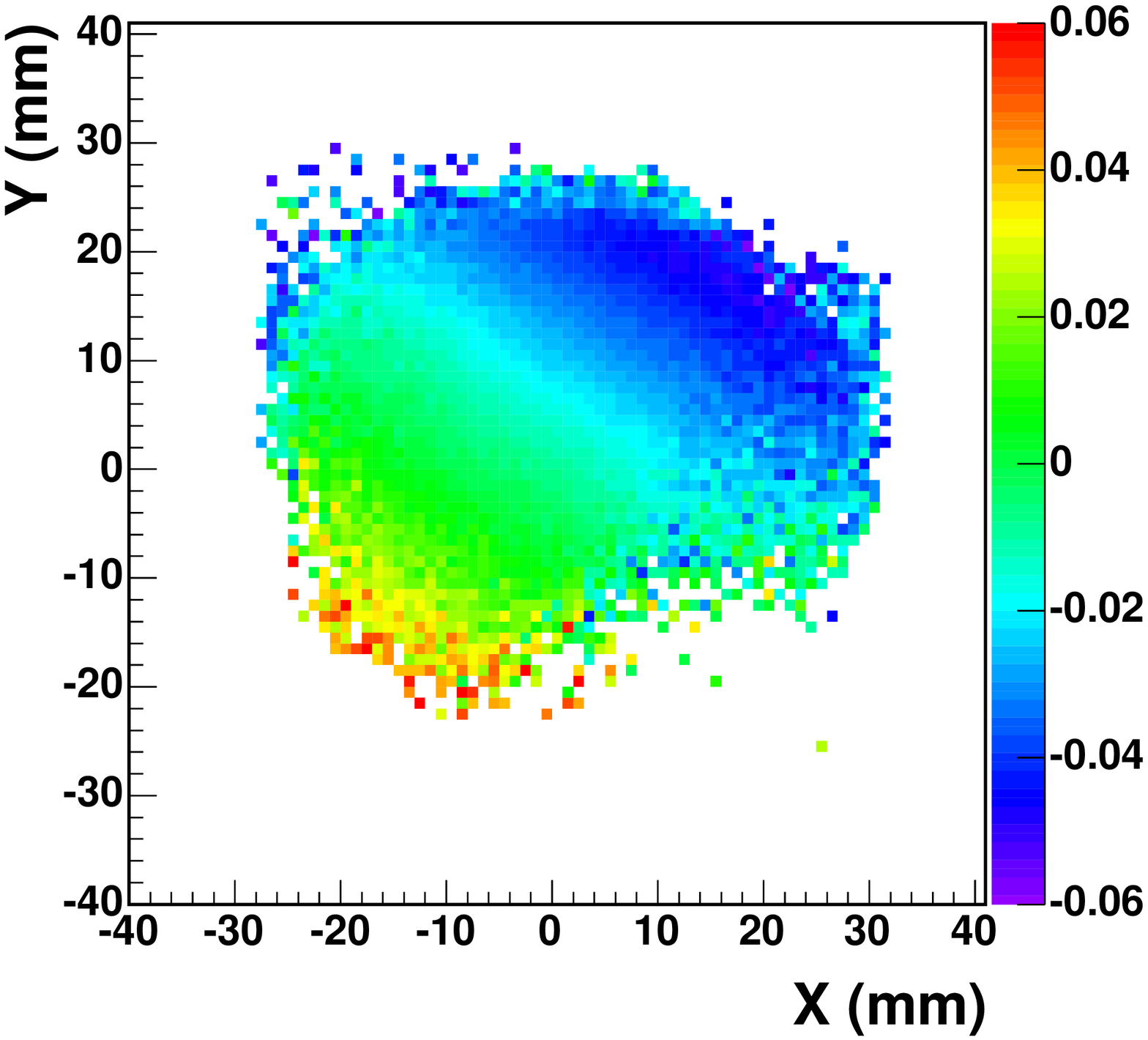}
\end{center}
\end{minipage}
     \caption{Two-dimensional distributions of the muon beam mean
       angle with respect to the $z$ axis in the $xz$ plane 
       ($\theta_x$, left) and in the $yz$ plane ($\theta_y$,
       right) for the beam intensity
       distribution shown in Fig. \ref{beam_yvsx}. Note that the
       color scales, representing the angles in radians, are
       different to show the trends more clearly. The plots show 
       the rotation of the
       angle-position correlation due to the fringe field of the
       solenoid, and the smaller divergence of the beam in the $x$
        direction than in $y$ for this particular beam tune.  }
     \label{dx_dy_yvsx}
\end{figure}

The inputs to the simulation are: the probability of a muon at
each $xy$ position as derived from a two-dimensional distribution
such as Fig. \ref{beam_yvsx}, the mean angles of the muon in $x$
($\bar{\theta_x}$) and $y$ ($\bar{\theta_y}$) for each $xy$
position as shown in Fig. \ref{dx_dy_yvsx}, and the rms of the
angle distributions for each $xy$ position. The simulation
generates an $xy$ transverse location according to the
probability of Fig. \ref{beam_yvsx}, using an array with elements
of $1 \times 1$~mm$^2$ as shown in the figures. Random Gaussian
numbers are generated with mean given by the values of
$\bar{\theta_x}$ and $\bar{\theta_y}$ at the $xy$ point as in
Fig. \ref{dx_dy_yvsx}.  The corresponding rms widths used for the
random numbers at each $xy$ point are also taken from the TEC
information.  However, they must be reduced compared to the
measured values, to account for the scattering added by the
materials of the TECs as the particle passes through; this is
accomplished by a quadratic subtraction, using scattering
contributions estimated from comparisons of simulations with
and without multiple scattering.  This technique is
intended to reproduce the correlations found between the angles,
$\theta_x$ and $\theta_y$, and the positions, $x$ and $y$.
Correlations between $\theta_x$ and $\theta_y$ have been observed
to be very small and are not simulated.

\begin{figure}[hbtp]
   \begin{center}
     \includegraphics[width=150mm]{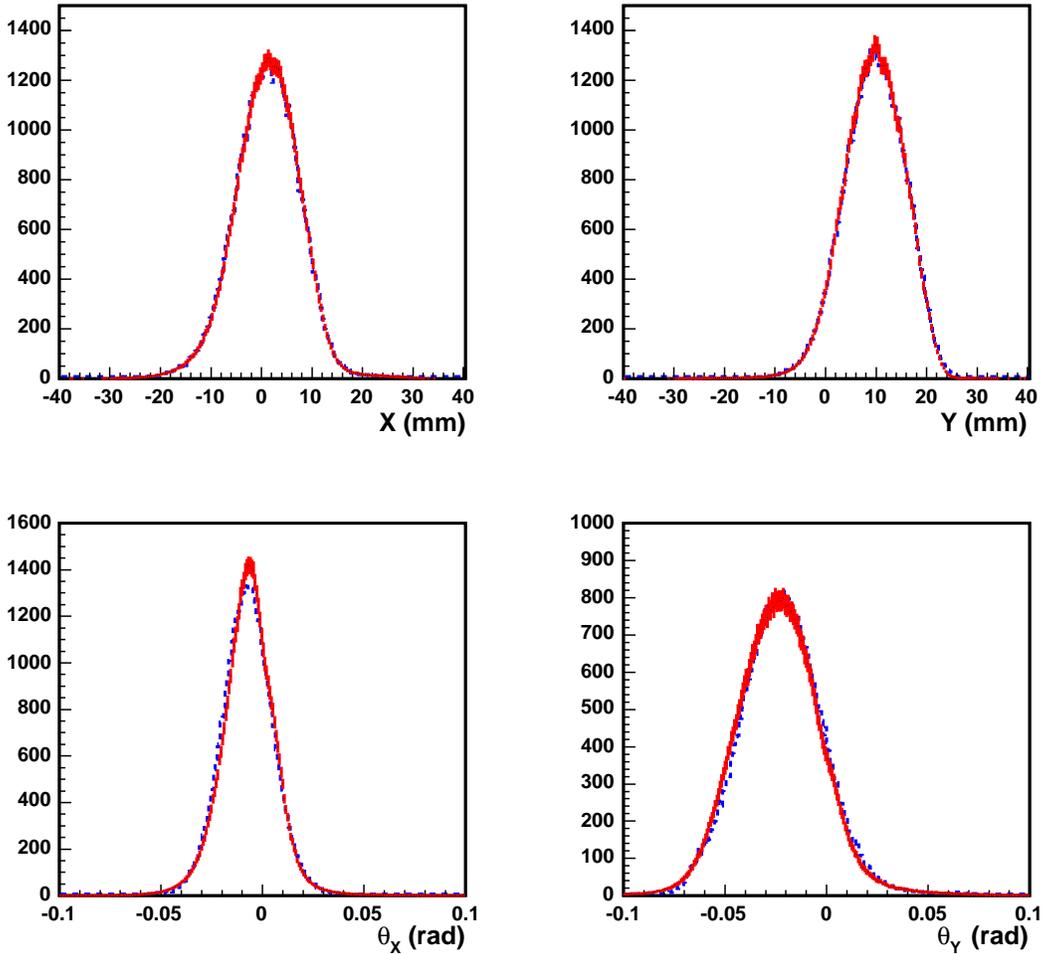}
     \caption{Distributions in positions and angles of a beam as
       measured (solid red lines) and as simulated following the
     quadratic subtraction procedure (dashed blue lines).}
     \label{beam_ms_deconvolution}
   \end{center}
\end{figure}

The muon is scattered by the TEC gas box entrance window ($2.1
\times 10^{-5}$ radiation lengths) and by 3.5~cm of DME gas ($1.3
\times 10^{-5}$ radiation lengths) before it reaches the TEC X
module, resulting in a smearing of the measured $\theta_x$ by an
estimated 9.5~mrad.  Scattering has an even larger influence on
the $\theta_y$, smearing it by 12.6~mrad, since the muon travels
through an additional 12.0~cm of DME ($4.6 \times 10^{-5}$
radiation lengths) before it reaches the Y module.

Projections of real measured positions and angles are shown in
Fig. \ref{beam_ms_deconvolution}, along with a comparison from a
similar analysis of TEC data produced by a simulation that uses
the correlated quadratic subtraction of the real distributions as
an input.  When the distribution of angles for the entire beam is
plotted, including this contribution from multiple scattering,
the lower two panes of Fig.~\ref{beam_ms_deconvolution} show a
12.4~mrad rms width for $\theta_x$ and a 20.1~mrad rms width for
$\theta_y$.

\section{Summary}

The TEC system has already provided TWIST with important
information about the muon beams used to make precise
measurements of muon decay as a test of the Standard Model.  In
the future, we expect to be able to improve the optimization of
beam characteristics (beam size and divergence) to provide the
smallest possible depolarization of the beam as it enters the
solenoidal field of the spectrometer. We will also be able to
determine what factors affect the muon beam, and by how much.
This in turn will enable us to minimize systematic effects of the
muon decay parameter determinations caused by variations in the
muon beam, and also to assess accurately the influence of
residual systematics. Finally, data from the TECs will be crucial
to a reliable simulation of the muon beam, which in turn will be
necessary for the detailed simulation of all aspects of the
experiment.  Because the extraction of muon decay parameters from
data depends on a high precision comparison to the simulation,
the success of the experiment ultimately will depend in part on
the data supplied by the TECs.

We acknowledge with gratitude the assistance of P. Bennett, S.
Chan, L. Ellstrom, B. Evans, and M. Goyette, as well as the
entire TWIST collaboration, especially R.E Mischke for his
comments and suggestions. TEC system construction and development
was supported by TRIUMF and the National Research Council of
Canada.  TWIST is supported by research grants from the National
Sciences and Engineering Research Council of Canada, by the US
DOE, and by the Russian Ministry of Science.

\appendix

\section{Gas flow control}
\label{gas_appendix}

In the TWIST TEC system, flammable DME chamber gas is separated
from the cyclotron vacuum by aluminized Mylar windows of 6 $\mu$m
thickness.  Consequently, major design features of the gas system
include interlocks and procedures to prevent differential
pressure from rupturing the 6 $\mu$m windows and to avoid
flammable mixtures of air and DME in the TECs.  To provide
redundancy, an independent bipolar differential pressure
transducer is used to monitor continuously the differential
pressure across the windows.  A programmable logic controller is
used to read the various pressure, flow and other status inputs
and implement the interlock logic. The system has three modes of
operation; TEC\_Out, Pump/Vent, and Normal Operation.

The TEC\_Out mode is used when the TEC gas box has been removed from the
vacuum box. Whether or not the TECs are in the vacuum box is
detected by a microswitch on the vacuum box mating flange.  In
this mode, all the TEC supply and exhaust valves are forced
closed, thus isolating the gas system. High voltage to the TECs is
disabled.  The control system's only active functions are to
control in a safe way the gate valve separating the TEC vacuum
box from the beam line, and the valve connecting the vacuum box
to its vacuum pump and vent ports, with interlocks based on the
pressures reported by vacuum gauges in the TEC vacuum box and the
beam line.

In the Pump/Vent mode, some valves are forced open, while others
are forced closed, such that the TEC gas box and all of the connecting
supply and exhaust tubing, which can be at sub-atmospheric
pressure during normal operation, can be pumped out through a
bypass valve connecting the TEC gas box to the TEC vacuum box.  This
ensures that possible air leaks into the TECs or any of the
connecting plumbing can be discovered.  The gate valve is forced
closed in this mode, isolating the TEC vacuum box from the beam
line during vent and pump down procedures.  High voltage to the
TECs is disabled.  The bypass valve connecting the TEC gas box to the
vacuum box can be closed to assist in locating leaks. However, it
is forced open if the differential pressure across the windows
exceeds a set point limit.

The Normal Operation mode has the most complex interlock scheme.
The bypass valve and the valve connecting the vacuum box to the
vent and vacuum pump ports cannot be opened.  The TEC high
voltages are enabled only if all the valves in the gas supply and
exhaust lines are open and the DME flow to the TECs exceeds a
set point limit (typ. 30 cc/min).  The gate valve can be open only
if the vacuum box and beam line pressures are satisfactory and
the differential pressure across the TEC gas box windows is below a first
set point limit.  The gas system valves in the DME supply and
exhaust lines can be open only if the vacuum box pressure is
satisfactory, and the differential pressure across the windows is
below the first set point limit.  If despite being isolated from
the gas system, the magnitude of the differential pressure
continues to increase beyond a second, higher set point limit, the
bypass valve connecting the TEC gas box to the vacuum box is forced open
to relieve the pressure across the windows.  This second limit is
set at $\pm 160$ mbar, well below the $\pm 500$~mbar typically
required to rupture the windows.


\end{document}